\documentclass{ws-ijmpe}
\usepackage[super,compress]{cite}
\usepackage[breaklinks]{hyperref}
\hypersetup{colorlinks,urlcolor=black,citecolor=black,linkcolor=black,filecolor=black}
\usepackage{breakurl}
\usepackage{braket}

\begin{document}

\markboth{K. J. Fushimi, M. M. Saez, O. Civitarese, M. E. Mosquera}{ANDES laboratory prospects}

\catchline{}{}{}{}{}

\title{Dark matter, supernova neutrinos and other backgrounds in direct dark matter searches. The ANDES laboratory prospects}

\author{K. J. Fushimi}

\address{Facultad de Ciencias Astron\'omicas y Geof\'{\i}sicas, University of La Plata,\\ Paseo del Bosque S/N
1900, La Plata, Argentina.\\
kfushimi@fcaglp.unlp.edu.ar}

\author{M. M. Saez}

\address{Facultad de Ciencias Astron\'omicas y Geof\'{\i}sicas, University of La Plata,\\ Paseo del Bosque S/N
1900, La Plata, Argentina.\\
msaez@fcaglp.unlp.edu.ar}

\author{M. E. Mosquera}

\address{Dept. of Physics, University of La Plata, \\ c.c.~67
 1900, La Plata, Argentina.\\Facultad de Ciencias Astron\'omicas y Geof\'{\i}sicas, University of La Plata,\\ Paseo del Bosque S/N
1900, La Plata, Argentina.\\
mmosquera@fcaglp.unlp.edu.ar}

\author{O. Civitarese}

\address{Dept. of Physics, University of La Plata, \\ c.c.~67
 1900, La Plata, Argentina.\\
osvaldo.civitarese@fisica.unlp.edu.ar}

\maketitle

\begin{history}
\received{Day Month Year}
\revised{Day Month Year}
\end{history}

\begin{abstract}
Dark Matter particles can be detected directly via their elastic scattering with nuclei. Next generation experiments can eventually find physical evidences about dark matter candidates. With this motivation in mind, we have calculated the expected signals of dark matter particles in xenon detectors. The calculations were performed by considering different masses and parameters within the minimal supersymmetric standard model. Since the detectors can also detect neutrinos, we have analyzed the supernova neutrino signal including a sterile neutrino in the formalism. Using this $3+1$ scheme we make predictions for both the normal and inverse mass hierarchy. In order to perform a study of the response of planned direct-detection experiments, to be located in ANDES (Agua Negra Deep Experimental Site), we have calculated the neutrino contributions to the background by taken into account reactor's neutrinos and geoneutrinos at the site of the lab. As a test detector, we take a Xenon1T-like array.
\end{abstract}


\ccode{PACS numbers:}

\section{Introduction and motivation:}
\label{sec:intro}

Several advances in experimental techniques to study large-scale structures have been achieved in recent years. Although dark matter has not been detected yet, and its properties remain largely unknown, there is evidence of its existence in astronomical observations of galaxy clusters and other cosmological observables \cite{Rubin:1970,zwicky:1933,Clowe:2006,Planck:2018,Schumann:2019}.\\

Cold dark matter candidates come from theories beyond the standard model. The most promising alternatives are axions and WIMPs (weakly interacting massive particles) \cite{majumdar:2014}. The notion about axions was introduced by Peccei and Quinn \cite{Peccei:1977} in connection with the so-called CP problem in QCD \cite{chadhaday:2021,Holman:1983}. The expected mass of the axion (scalar,boson) is of the order of $2-6\times10^{-9}$ eV \cite{Perez:2020}; however, if non-linear observables are considered, this value may be much lower ($ 10^{-22}$ eV) \cite{Marsh:2016}. The technique to detect them experimentally consists of the search for photons generated by their interactions with axions through the Primakoff mechanism \cite{Freese:2017}.\\

WIMPs are believed to be decoupled in the early Universe in thermal equilibrium and are expected to have masses between $ 1-1000 $~GeV \cite{Gelmini:2017}. The currently most accepted dark matter candidate is the lightest neutralino \cite{Jungman:1996} which is a linear combination of supersymmetric particles. Direct, indirect and particle collider methods can be used to detect them experimentally \cite{Gelmini:2016thdm}. This work focuses only on WIMPs since there is still a region in the cross-section vs mass plane allowed by the current experimental constraints.\\

Direct detection experiments are today one of the techniques with the highest expectations for detecting dark matter particles. The scattering of WIMPs by nuclei is usually divided into two channels: i) scalar type or spin independent; and ii) axial-vector type or spin dependent. In the first case, the cross-section increases with the mass of the nucleus, while in the second, the cross-section is proportional to $ J (J + 1) $, where $ J $ is the total angular momentum of the nucleus \cite{Bertone:2005}. If there is a halo composed of WIMPs in the galaxy, many of them should pass through the Earth and interact with matter. In order to detect these particles, it is possible to study the recoil energy of nuclei produced by the dispersion of WIMPs \cite{Goodman:1985, Wasserman:1986}. Essential ingredients for signal calculation in direct detection experiments are the density and velocity distribution of the WIMPs in the vicinity of the Sun and the WIMP-nucleus scattering cross-section \cite{Bertone:2005}. The reaction rate in direct detection experiments undergoes annual and diurnal modulations as a consequence of the annual revolution of the Earth around the Sun and the rotation of the Earth on its axis. If a modulation in the flow of these particles were observed on Earth, some properties of cold dark matter may be determined \cite{Freese:2012}. The DAMA collaboration claims to have observed the annual modulation in the energy range between $ 1-6 $ ~ keV \cite{Bernabei:2018}. Other collaborations such as Xenon1T \cite{XENON1T:2018} and LUX \cite{LUX:2017_1}, have reported null results.

These experiments eventually reach a background (called neutrino floor) due to the coherent scattering of neutrinos produced from different sources (Sun, atmosphere, Earth, and reactors). The elastic scattering of neutrinos on protons and nuclei (CE$\nu$NS) is an alternative tool to detect astrophysical neutrinos in the direct detection experiments \cite{Beacom:2002, Drukier:1984}. In particular, supernova neutrinos with energies of $\sim 10-20$ MeV can induce nuclear recoils through coherent elastic neutrino-nucleus scattering at the keV scale. This interaction is proportional to the square of the neutron number of the nuclei \cite{Lang:2016}. 

Core-collapse supernovae (SN) are the final evolutionary stage of stars with masses heavier than 8 $M_\odot$ and represents a long-awaited observation target for neutrino telescopes. Neutrinos are a key piece for studying SN, since all flavors are produced, and they travel through stellar material and space to reach a detector on Earth, giving information from deep inside the stellar core. The analysis and reconstruction of SN neutrino fluxes is an interesting tool to clarify the role of neutrinos in stellar explosion events and nucleosynthesis \cite{Qian:2018}, as well as for studying physics at high densities and the neutrino oscillation phenomena \cite{Mirizzi:2016,Takahashi:2001,Balantekin:2004, Machado:2012, Penacchioni:2019,Penacchioni:2019V2,Penacchioni:2020}. 

The possibility of the existence of a light sterile neutrino state is motivated by anomalies detected in short-baseline neutrino oscillation experiments \cite{Aguilar-Arevalo:2018,Dentler:2018}, in reactor experiments \cite{Mention11} and Gallium detectors \cite{Acero:2007, Giunti:2011}. This extra neutrino does not participate in weak interaction processes, and only interacts gravitationally, however it can participate in the mixing processes with active neutrinos \cite{Kopp:2013}. The consequences of the existence of sterile neutrinos in different astrophysical scenarios were already examined \cite{Boyarsky:2009,Mohapatra:2004}. In the context of supernovae, several authors have already analysed the effects of the inclusion of a sterile flavor and its oscillations upon the fraction of free neutrons, the baryonic density, the electron fraction of the material, and nucleosynthesis processes \cite{Fetter:2003,Qian:2018,Saez:2018,Tamborra:2012,Wu:2014,Esmaili:2014,Tang:2020}.

At present, there are several operationally dark matter detectors \cite{Bernabei:2018, CRESST:2019, CDMSlite:2019, Xenon1T:2019, SABRE:2019, Mimac:2013, COSINE100:2019, DarkSide:2018, PICO60:2019, Ajaj:2019}, which should also be able to detect neutrinos from the next galactic explosion. The currently running detectors are located in the Northern Hemisphere, however there are two projects to install detectors in the South. One is the SABRE experiment in Australia \cite{SABRE:2019}, whose goal is to build a twin detector to the one used by DAMA. The other is the ANDES Laboratory \cite{andesweb, Civitarese:2015}, which consists of the design and construction of an underground laboratory adjacent to the Agua Negra Tunnel complex in San Juan, Argentina. 

In line with previous studies focused on the ANDES laboratory \cite{Machado:2012, Civitarese:2016}, here we present a complementary study on the expected signals of dark matter, SN neutrinos, and the neutrino floor, in order to have a better understanding of the expected background at the site. In addition, we compare the signals calculated for ANDES with those of the Xenon1T detector \cite{Xe1T:2020}, which has fixed the most significant constraints.

The paper is organized as follows. In Section \ref{formalismo_wimps} we present a brief description of the dark matter model and the formalism needed to compute the direct detection rate. In Section \ref{formalismo-SN} we focus on the SN signal properties, and in Section \ref{andes_formalism} we derive the expression of the cross section and introduce the parameters used for a test detector to be located at the ANDES laboratory. In Section \ref{results} we show and discuss the results of the calculations. The conclusions are drawn in Section \ref{conclusion}.\\

\section{Dark matter direct detection}\label{formalismo_wimps}
\subsection{Dark matter model}

We work in the framework of the MSSM (Minimal Supersymmetric Standard Model). The lightest neutralino state can be written as a linear combination of binos ($\tilde{B}$), winos ($\tilde{W_3}$) and higgsinos ($\tilde{H}_1,\tilde{H}_2$) \cite{Engel:1992}
 \begin{equation}
      \chi_1^0=Z_{11}\tilde{B}+Z_{12}\tilde{W_3}+Z_{13}\tilde{H}_1+Z_{14}\tilde{H}_2\,.
      \label{chi}
 \end{equation}
In the simplest model, the coefficients of the linear combination ($Z_{11},\, Z_{12},\, Z_{13},\, Z_{14}$ ) depend on four SUSY parameters, the bino and wino mass parameters ($M'$ and $M$), the higgsino mass parameter ($\mu$), and $\tan{\beta}$ which is the ratio of vacuum expectation values of the two Higgs scalars. In the Grand-Unified-Theory (GUT), the parameters $M$ and $M'$ are related by $M' = \frac{5}{3} M \tan^2{\theta_W}$ \cite{Murakami:2001,Ellis:2000,Cerdeno:2001}.
 The effective lagrangian density describing the neutralino-quark elastic scattering in the MSSM is written \cite{Engel:1992}
\begin{equation}
 \small{
  \mathcal{L}_{eff}=\frac{g^2}{2M_W^2}\sum_q\left(\bar{\chi}\gamma^{\mu}\gamma_5\chi\bar{\psi_q}\gamma_{\mu}A_q\gamma_5\psi_q+\bar{\chi}\chi S_q \frac{m_q}{M_W}\bar{\psi_q}\psi_q\right)\,,}
\label{lagrangiano}
\end{equation}
where $g$ is the SU(2) coupling constant, $M_W$ stands for the mass of the W boson and $A_q$ and $ S_q$ are defined by \cite{Engel:1992}
\begin{eqnarray}
  A_q&=&\frac{1}{2} T_{3q}(Z^2_{13}-Z^2_{14})-\frac{M^2_W}{M^2_{\tilde{q}}}\Bigg(\left[T_{3q}Z_{12}-(T_{3q}-e_q)Z_{11}\tan{\theta_W}\right]^2\nonumber\\
  &&\hspace{4.4cm}+e_q^2Z_{11}^2\tan^2{\theta_W}+\frac{2m_q^2d_q^2}{4M_W^2}\Bigg)\,,
  \label{A_q}
\end{eqnarray}
\begin{eqnarray}
S_q&=&\frac{1}{2} \left(Z_{12}-Z_{11}\tan{\theta_W}\right)\Bigg[\frac{M^2_W}{M^2_{H_2}}g_{H_2}k_q^{(2)}+\frac{M^2_W}{M^2_{H_1}}g_{H_1}k_q^{(1)}+\frac{M^2_W \epsilon d_q}{M^2_{\tilde{q}}}\Bigg]\,,
\end{eqnarray}
where $M_{\tilde{q}}$ and $M_{H^0_i}$ are the squark and higgsino
masses respectively \cite{Djouadi:2008,Pdg:2019}, $T_{3q}$,
$e_q$, and $m_q$ are the quark weak isospin, charge and mass
respectively, $\epsilon$ is the sign of the lightest-neutralino mass
eigenvalue \cite{Engel:1992}. The $d_q$ and $k_q^{(i)}$ parameters for the up-type and down-type quark are taken from reference \cite{Djouadi:2008}.

We compute the cross-section using the Lagrangian of Eq.~\eqref{lagrangiano}. For more details on the calculation of the cross-section see ref. \cite{Fushimi:2020} and references therein.
\subsection{Detection rates}

The differential recoil rate per unit mass of the detector can be defined as \cite{Freese:2012}
\begin{eqnarray}
\label{rate2}
\frac{dR}{dE_{\rm nr}}&=&\frac{2\rho_{\chi}}{m_{\chi}} \frac{\sigma_0}{4 \mu^2 } F^2(q) \eta\, ,
\end{eqnarray}
in units of $\textrm{cpd}\textrm{kg}^{-1} \textrm{keV}^{-1}$, where cpd stands for counts per day. In the previous equation $E_{\rm nr}$ is the nuclear recoil energy,  $q^2= 2 m_A E_{\rm nr}$, $m_A$ is the mass of the nucleus, $m_{\chi}$ and $\rho_{\chi}$ are the dark matter mass and the local density, $\rho_{\chi}=0.3$~GeV/cm$^3$ \cite{Schumann:2019,Gelmini:2017}, $\sigma_0$ is the cross-section at $q=0$ and $F(q)$ stands for the nuclear form factor. We define the mean inverse-velocity as
\begin{eqnarray}
\label{inv-vel}
\eta&=&\int{\frac{f(\vec{v},t)}{v}d^3v} \, ,
\end{eqnarray}
where $f(\vec{v},t)$ is the WIMP velocity distribution and $v$ is the speed of the WIMP relative to the nucleus.

\subsubsection{Velocity distribution}

For the velocity distribution of the WIMP, we assume that the model for the DM halo is the Standard Halo Model \cite{Freese:1988}; therefore, we calculate the velocity distribution from the truncated Maxwell-Boltzmann distribution \cite{Freese:2012,Jungman:1996}
\begin{eqnarray}
f({\vec{v}})&=&\left\{
\begin{array}{cc}
\frac{1}{N(\pi v_0^2)^{\frac{3}{2}}}e^{{-\left|{\vec{v}}\right|^2}/{v_0^2}} & |\vec{v}|<v_{\rm esc} \\
0&|\vec{v}|>v_{\rm esc} \\
\end{array}
\right. \, , 
\end{eqnarray}
where $N$ is a normalization factor 
and $v_{\rm esc}$ and $v_0$ are the escape velocity and the velocity of the Sun, respectively, and their values are $v_{\rm esc}=544 \, {\rm km/s}$ \cite{Gelmini:2015} and $v_0=220 \, {\rm km/s}$ \cite{Jungman:1996}. If one considers the laboratory-velocity $\left(\vec{v}_{\rm lab}\right)$, the integral of Eq.~\eqref{inv-vel} becomes
\begin{eqnarray}
\label{eta1-2}
\eta&=&\frac{1}{N(\pi {v_0}^2)^{3/2}}\int{ \frac{e^{{-({\vec{v}}+{\vec{v}_{\rm lab}})^2}/{v_0^2}}}{v}d^3v}\, .
\end{eqnarray}

\subsubsection{Cross-section}
The total WIMP-nucleus cross-section is a sum over the spin-independent (SI) and spin-dependent (SD) contributions. Since, in this work, we are dealing with the spin independent channel, the differential cross-section is written \cite{Freese:2012}:
\begin{align}
  \label{dif-seccioneficaz-nucleo}
  \frac{d\sigma}{dE_{\rm nr}}(E_{\rm nr},v)=& \frac{m_A}{2v^2\mu_{A}^2}\sigma_{SI}(0)F_{SI}^2(E_{\rm nr})\,,
\end{align}
where $\mu_{A}$ is the reduced mass of the WIMP-nucleus system. For this cross-section we consider all the mediators that contribute to the scalar interaction, that is, $ H_1 $, $ H_2 $ and $ \tilde{q} $, and a Helm nuclear form factor \cite{Helm:1956}
\begin{equation}
  F(q)=3 e^{-q^2s^2/2} \frac{\sin{(qr_n)}-qr_n\cos{(qr_n)}}{(qr_n)^3} \,,
  \label{Helm}
\end{equation}
where $s \simeq 0.9$ fm, $r_n^2=c^2+\frac{7}{3}\pi^2a^2-5s^2$ is an effective nuclear radius, $a \simeq 0.52$ fm and $c \simeq (1.23 A^{1/3}- 0.6 )$ fm \cite{Lewin:1995}.

\subsection{Annual and diurnal modulation signal}
The direct detection experiments aim to detect modulations in the signal. This phenomenon is the result of the revolution of the Earth around the Sun.
The rate can be expanded in a Taylor series as \cite{Civitarese:2016}
\begin{align}
    \frac{dR}{dE_{\rm nr}} \simeq &\Bigg\{ S_0+S_{\rm m}(E_{\rm nr})\cos(w_{\rm rev}(t-t_{\rm a}))+S_{\rm d}(E_{\rm nr})\cos(w_{\rm rot}(t'-t'_{\rm d}))\Bigg\}\,,
\end{align}

where $t_{\rm a}= t_{\rm eq}+\frac{\beta_{\rm m}}{w_{\rm rev}}$, $t'_{\rm d}=\frac{\beta_{\rm d}}{w_{\rm rot}}-t_{0}$, we define $w_{\rm rev}$ as the Earth revolution frequency, $t_{\rm eq}$ is the sidereal time corresponding to the March equinox (that is the sidereal day 80.22 referred to J2000.0), $\beta_{\rm m}=1.260$ rad, $\beta_{\rm d}=3.907$ rad, $w_{\rm rot}=2\pi/ (\textrm{1 sidereal day})$, and $t_{0}$ is the time corresponding to the longitude of the laboratory $\lambda_0$. For more details on the dependence of the recoil rate with the laboratory coordinates see Reference \cite{Civitarese:2016}.

In the previous equation $S_0$ is the time-average of the rate, which does not present a modulation. $S_{\rm m}$ is the annual modulation amplitude and $S_{\rm d}$ is the diurnal modulation amplitude.

To compare with experimental data, we integrate the modulation amplitude on an interval of the bin energy that is defined by the resolution of the detector \cite{Savage:2009,Freese:2012,Civitarese:2016}. Since the detectors do not measure the nuclear recoil energy directly, it is necessary to include a quenching factor $Q$, which relates the electron equivalent energy $E_{ee}$ measured by the detector with the recoil energy ($E_ {ee}=Q E_ {nr}$) \cite{Freese:2012}, therefore

\begin{align}
\langle S_{\rm m}\rangle=&\frac{1}{\Delta E}\int_{E_1}^{E_2} S_{\rm m}(E_{\rm nr}) \varepsilon(E_{\rm nr}Q) \Phi(E_{\rm nr}Q,E_1,E_2)dE_{\rm nr} \, ,\\
\label{sdmedia}
\langle S_{\rm d}\rangle=&\frac{1}{\Delta E}\int_{E_1}^{E_2} S_{\rm d}(E_{\rm nr})\varepsilon(E_{\rm nr}Q) \Phi(E_{\rm nr}Q,E_1,E_2)dE_{\rm nr}\, .
\end{align}
where $\Delta E= E_2-E_1$ is the bin length, $\varepsilon(E_{\rm nr}Q)$ is the efficiency of the experiment and $\Phi(E_{ee},E_1,E_2)$ is a response function corresponding to the fraction of events with an expected observed energy \cite{Savage:2009}.

\section{Supernova neutrino emission and detection}\label{formalismo-SN}
\subsection{SN neutrino fluxes}

To model the SN explosion, we consider the standard energy released by the supernova neutrino outflow to be similar to the SN1987A \cite{Hirata:1987,Bionta:1987}, that is $E_\nu^{tot}=3\times 10^{53} \, {\rm erg}$ distributed on the different neutrino's flavors. The luminosity flux for each flavor is time dependent and can be written as a function of the post-bounce time ($t_{pb}$) as $L_{\nu_\beta}(t_{pb})=\frac{E_\nu^{tot}}{18} e^{-t_{pb}/3}$ \cite{Huang:2015}.

To characterize the neutrinos emerging from the proto-neutron star, we use a distribution function based on the following power-law parametrization proposed by the Garching group \cite{Keil:2003,Tamborra:2012}, valid for electron, muon, and tau neutrinos
\begin{equation}
f_{\nu_\beta}(E)= \frac{{(\kappa+1)}^{(\kappa+1)}}{\Gamma(\kappa+1)\braket{E}} \left(\frac{E}{\braket{E}}\right)^\kappa e^{-(\kappa+1)E/\braket{E}}\,  ,
\label{eq:PL}
\end{equation}
where $E$ is the neutrino energy. Along this work, we have fixed $\kappa=3$ for all flavors and mean energies $\braket{E_{\nu_e}}=12 \, {\rm MeV}$, $\braket{E_{\bar{\nu}_e}}=15 \, {\rm MeV}$ and $\braket{E_{\nu_x}}=\braket{E_{\bar{\nu}_x}}=18 \, {\rm MeV}$ \cite{Machado:2012,2010:Hudepohl}. For the case of sterile neutrinos the Eq.\eqref{eq:PL} is multiplied by a hindrance factor $h_s$ ($0\le h_s \le 1$), see section \ref{SNanalysis}.
The spectral flux, for each neutrino flavor produced in the SN explosion to be detected at a distance $D$, is given by
\begin{equation}\label{eq:ini-fluxes}
F^0_{\nu_\beta}(E,\, t_{pb})=\frac{L_{\nu_\beta}(t_{pb})}{4\pi D^2} \frac{ f_{\nu_\beta}(E)}{\braket{E_{\nu_\beta}}}\,\,\, ,
\end{equation}
where $\braket{E_{\nu_\beta}}$ is the mean energy of the $\beta$-flavor neutrino eigenstate. 

Direct dark matter detectors are sensitive to all neutrino flavors. Thus the oscillations between active flavors would not generate any effect with respect to the case without oscillations, because the total flux is conserved. We shall refer to this case, all along the text, as the {\it {standard}} case. But, if active-sterile oscillations are considered, distinctive features may appear in the detected signals.  We shall refer to this case as the $3+1$-scheme.  Following \cite{Esmaili:2014}, the fluxes in the $3+1$-scheme can be written as
{\small
\begin{align}\label{flujos-3+1}
F_{\nu_e}&= \Theta_e^e F^0_{\nu_e} + \Theta_e^x F^0_{\nu_x} + \Theta_e^s  F^0_{\nu_s}  \, \, \, , \nonumber \\
F_{\bar{\nu}_e}&= \Xi_e^e F^0_{\bar{\nu}_e}+ \Xi_e^x F^0_{\bar{\nu}_x}+\Xi_e^s F^0_{\bar{\nu}_s}  \, \, \, , \nonumber \\
F_{\nu_x}&= \left(\Theta_\mu^e +\Theta_\tau^e\right) F^0_{\nu_e} + \left(\Theta_\mu^x +\Theta_\tau^x\right) F^0_{\nu_x} + \left(\Theta_\mu^s +\Theta_\tau^s\right)  F^0_{\nu_s}\, \, \, , \nonumber\\ 
F_{\bar{\nu}_x}&= \left(\Xi_\mu^e +\Xi_\tau^e \right) F^0_{\bar{\nu}_e}+ \left(\Xi_\mu^s +\Xi_\tau^s \right)F^0_{\bar{\nu}_x}+ \left(\Xi_\mu^s +\Xi_\tau^s \right) F^0_{\bar{\nu}_s}\, \, \, , \nonumber\\
F_{\nu_s}&=\Theta_s^e F^0_{\nu_e} + \Theta_s^x F^0_{\nu_x} +\Theta_s^s F^0_{\nu_s}\, \, \, , \nonumber \\
F_{\bar{\nu}_s}&=\Xi_s^e F^0_{\bar{\nu}_e}+\Xi_s^x  F^0_{\bar{\nu}_x}+\Xi_s^s F^0_{\bar{\nu}_s} \, \, \, ,
\end{align}}
where for normal hierarchy we have defined
{\small
\begin{align}
\Theta_{\alpha}^e=& |U_{\alpha 1}|^2P_H P_L (1-P_S)+|U_{\alpha 2}|^2 P_H(1-P_L)(1-P_S) \nonumber \\
& +|U_{\alpha 3}|^2P_S+|U_{\alpha 4}|^2(1-P_S)(1-P_H) \, \, \, ,\nonumber\\
\Theta_{\alpha}^x=& |U_{\alpha 1}|^2(1-P_H P_L) +|U_{\alpha 2}|^2(1-P_H+P_H P_L)+|U_{\alpha 4}|^2P_H  ,\nonumber\\
\Theta_{\alpha}^s=&|U_{\alpha 1}|^2P_S P_H P_L+|U_{\alpha 2}|^2P_S P_H(1-P_L) \, \, \, , \nonumber \\
& +|U_{\alpha 3}|^2(1-P_S)+|U_{\alpha 4}|^2P_S(1-P_H)  \, \, \, , \nonumber \\
\Xi_{\alpha}^e=&|U_{\alpha 1}|^2\, \, \, , \nonumber \\
\Xi_{\alpha}^x=& |U_{\alpha 2}|^2+|U_{\alpha 3}|^2\, \, \, , \nonumber \\
\Xi_{\alpha}^s=& |U_{\alpha 4}|^2\, \, \, ,
\end{align}}
and for the inverse hierarchy we have 
{\small
\begin{align}
\Theta_{\alpha}^e=& |U_{\alpha 1}|^2P_L (1-P_S)+|U_{\alpha 2}|^2P_S \nonumber\\
&+|U_{\alpha 4}|^2(1-P_S)(1-P_L) \, \, \, ,\nonumber\\
\Theta_{\alpha}^x=& |U_{\alpha 1}|^2(1-P_L) +|U_{\alpha 3}|^2+|U_{\alpha 4}|^2 P_L  \, \, \, ,\nonumber\\
\Theta_{\alpha}^s=&  |U_{\alpha 1}|^2P_SP_L+|U_{\alpha 2}|^2(1-P_S)+|U_{\alpha 4}|^2P_S(1-P_L) \, \, \, , \nonumber \\
\Xi_{\alpha}^e=& |U_{\alpha 2}|^2\bar{P}_H+|U_{\alpha 3}|^2(1-\bar{P}_H)\, \, \, , \nonumber \\
\Xi_{\alpha}^x=& |U_{\alpha 1}|^2+|U_{\alpha 2}|^2(1-\bar{P}_H)+|U_{\alpha 3}|^2\bar{P}_H\, \, \, , \nonumber \\
\Xi_{\alpha}^s=& |U_{\alpha 4}|^2\, \,. 
\end{align}}
In the last expressions $U_{\alpha k}$ are the elements of the neutrino mass-mixing matrix in the 3+1 scheme, for which we have adopted the same parametrization as in Reference \cite{Collin:2016} with $\theta_{14}\neq 0$ and $\theta_{24}=\theta_{34}=0$. $P_H$ ($\bar{P}_H$) and $P_L$ ($\bar{P}_L$) are the neutrino (antineutrino) crossing probabilities for the H and L resonances respectively, and $P_S$ stands for the probability of crossing the S-resonance (see refs. \cite{Esmaili:2014,Tang:2020} for details).

\subsection{Neutrino interactions in dark matter detectors}

The range of energy in which the detectors work allows the measurement of neutrino scattering when there are more than a tonne of material in the detector \cite{Kozynets:2019,raj:2020,Dutta:2019}.

Coherent elastic neutrino-nucleus scattering is mediated by the $Z_0$ boson and is sensitive to all neutrino flavors. The interaction of a SN neutrino with a Xenon nucleus in a detector through CE$\nu$NS causes the nucleus to recoil with a differential rate \cite{Lang:2016}
\begin{equation}\label{rate-SN}
    \frac{dR}{dE_{\rm nr}}=\sum_{\nu_\beta} N_{A} \int_{E_{min}} dE F_{\nu_\beta} \frac{d\sigma}{dE_{\rm nr}}(E,E_{\rm nr})\,,
\end{equation}
where the sum is performed over the neutrino flavors, and $N_{A}$ is the amount of target nuclei per tonne of material. $E_{min}=\sqrt{m_A E_{\rm nr}/2}$, with $m_A$ being the mass of the target nucleus. $F_{\nu_\beta}$ is the neutrino flux of $\beta$-flavor, and $\frac{d\sigma}{dE_{\rm nr}}$ is the corresponding neutrino-nucleus scattering cross-section \cite{Drukier:1984} 
\begin{equation}
    \frac{d\sigma}{dE_{\rm nr}}(E,E_{\rm nr})=\frac{G_F^2 m_A}{4\pi}Q^2_W\left(1-\frac{m_A E_{\rm nr}}{2E^2} \right) F^2(E_{\rm nr})\,,
\end{equation}
where $G_F$ is the Fermi constant, $Q_W=N-(1-4\sin^2(\theta_W))Z$, is the weak hypercharge ($N$ and $Z$ are the neutron and proton number respectively), and $F(E_{\rm nr})$ is the Helm nuclear form factor (see Eq.\eqref{Helm}).
\section{Counts and detection}\label{andes_formalism}

The total number of counts in the detector can be calculated as
\begin{equation} \label{counts_Xe1T}
N=\frac{1}{\Delta E}\int_{E_1}^{E_2} \frac{dR}{dE_{\rm nr}}(E_{\rm nr})\varepsilon(E_{\rm nr}Q) \Phi(E_{\rm nr}Q,E_1,E_2) dE_{\rm nr}\, ,
\end{equation}
where $\Delta E= E_2-E_1$ defines the energy interval, and $\frac{dR}{dE_{\rm nr}}$ are the differential recoil rates for WIMPs or SN neutrinos. In the calculations we have taken an equally-spaced bin distribution of 0.5 keV. The fraction of events with expected observed energy $E_{ee}$ is given by the response function $\Phi(E_{ee},E_1,E_2)$ which is written as \cite{Savage:2009} 
\begin{equation}
\label{phi}
    \Phi(E_{ee},E_1,E_2)=\frac{1}{2}\left[{\rm{erf}}\left(\frac{E_2-E_{ee}}{\sqrt{2}\sigma(E_{ee})}\right)-{\rm{erf}}\left(\frac{E_1-E_{ee}}{\sqrt{2}\sigma(E_{ee})}\right)\right]\,.
\end{equation}

To analyze the signal dependence upon the characteristics and location of the future ANDES laboratory, we consider a detector similar to that of Xenon1T experiment (unless otherwise specified). In Table~\ref{ANDES-parameters} we show the parameters corresponding to the ANDES site and our reference detector.

\begin{table}[ht!]
\begin{center}
\caption{Coordinates of the location of the ANDES laboratory and characterization of the potential Xenon test detector. The parameters entering in Eq.\eqref{phi} are given in the table together with the quenching factor $Q$, the efficiency $\varepsilon(E_{ee})$, the attenuation $\lambda=26.2$ and the threshold energy $E_{ee}^0=1.55$ keV \cite{Foot:2020,Xe1T:2020}.}
\begin{tabular}{cc}
\hline\hline
Location & $30^\circ 30'S$ $69^\circ53'W$\\ \hline
Target material&$^{131}$Xe (Z=54, N=77)\\
Q&1 \\
 $\varepsilon(E_{ee})$ & $0.87{\left(1+e^{-\lambda(E_{ee}-E_{ee}^0)}\right)}^{-1} $ \\
  $\sigma(E_{ee})$ & $(0.31 \rm{keV})\sqrt{E_{ee}/\rm{keV}}+0.0037E_{ee}$\\
\hline\hline
\end{tabular}
\end{center}

\label{ANDES-parameters}
\end{table}

\section{Results and discussion}\label{results}

\subsection{Analysis of dark matter signal}
We have computed the signal in the detector for different values of the neutralino mass and with the parameters entering Eq.~\eqref{rate2}. The masses for the squark and the higgs pseudo scalar ($M_{\tilde{q}}=1500$ GeV, $m_{ps}=500$ GeV) were extracted from ref. \cite{Pdg:2019}. We fixed the parameter $\tan{\beta}=10$ \cite{Ellis:2000,Cerdeno:2001,Murakami:2001}, and varied the parameter $\mu$. The value of the $M$ parameter was determined as a function of $\mu$ and $m_{\chi}$.

\begin{figure}[ht!]
\centering
\includegraphics[width=0.8\textwidth]{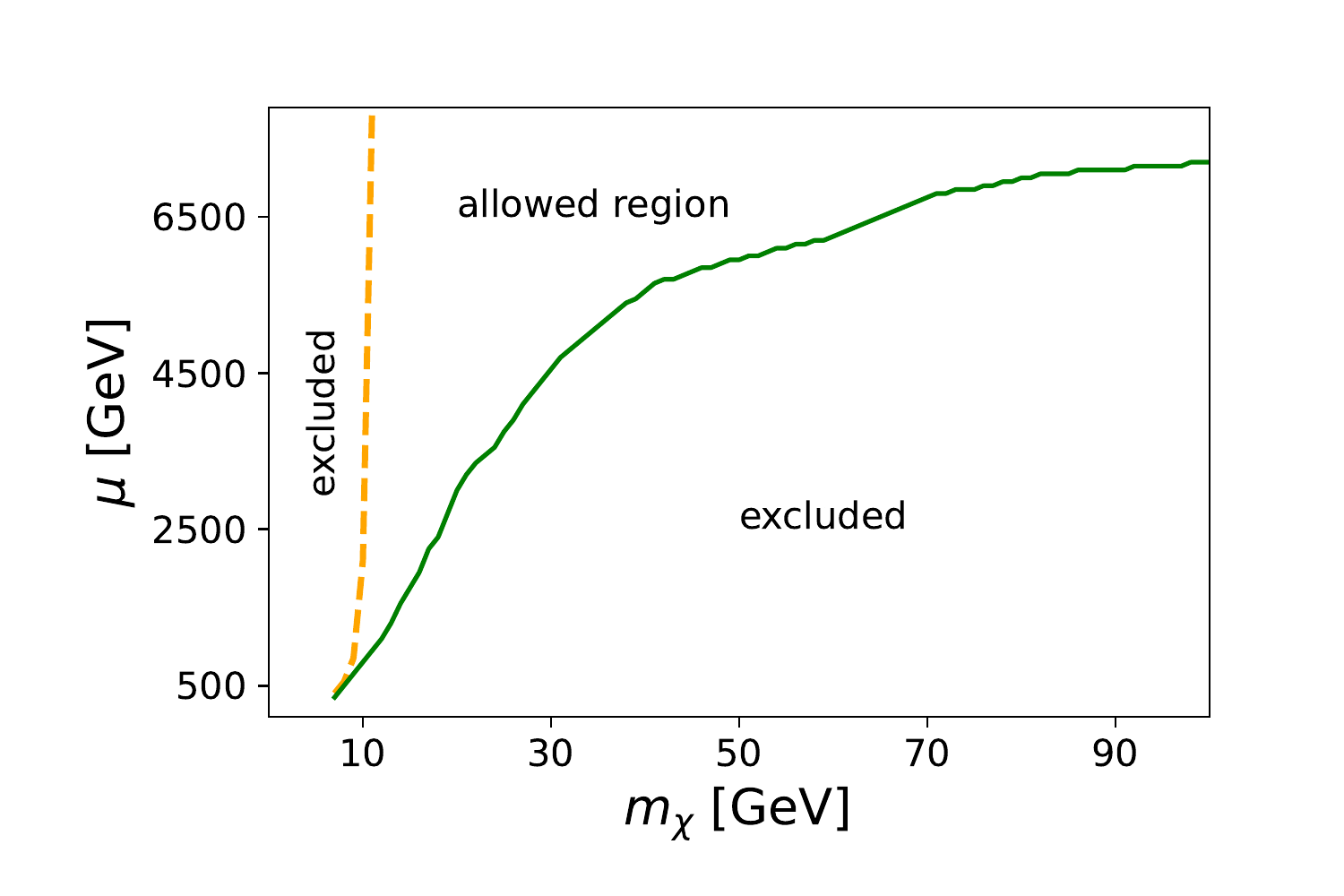}
\caption{$\mu$ and $m_{\chi}$ parameter space. The limits are the neutrino floor (dashed line) and the Xenon1T exclusion limit (solid line).}
\label{cotas_mu}
\end{figure}

The mass of the neutralino was calculated by diagonalizing its mass matrix.  In Figure ~\ref{cotas_mu} we present the explored parameter space, for $\mu$ and $m_{\chi}$  \cite{Fushimi:2020} for details. This region satisfies the limits imposed on the cross-section mass plane \cite{PDG:2020} given by the Xenon1T exclusion limit \cite{Xenon1T:2019}. In addition, we are looking for a signal that can be distinguished from the neutrino floor \cite{Schumann:2019}; this means that it should fall above that limit. Because the cross-section becomes lower the higher the value of $\mu$, the Xenon1T data gives a lower bound on this parameter, while the neutrino floor gives an upper bound, as shown in Figure \ref{cotas_mu}.

\begin{figure}[ht!]
\centering
\includegraphics[width=0.8\textwidth]{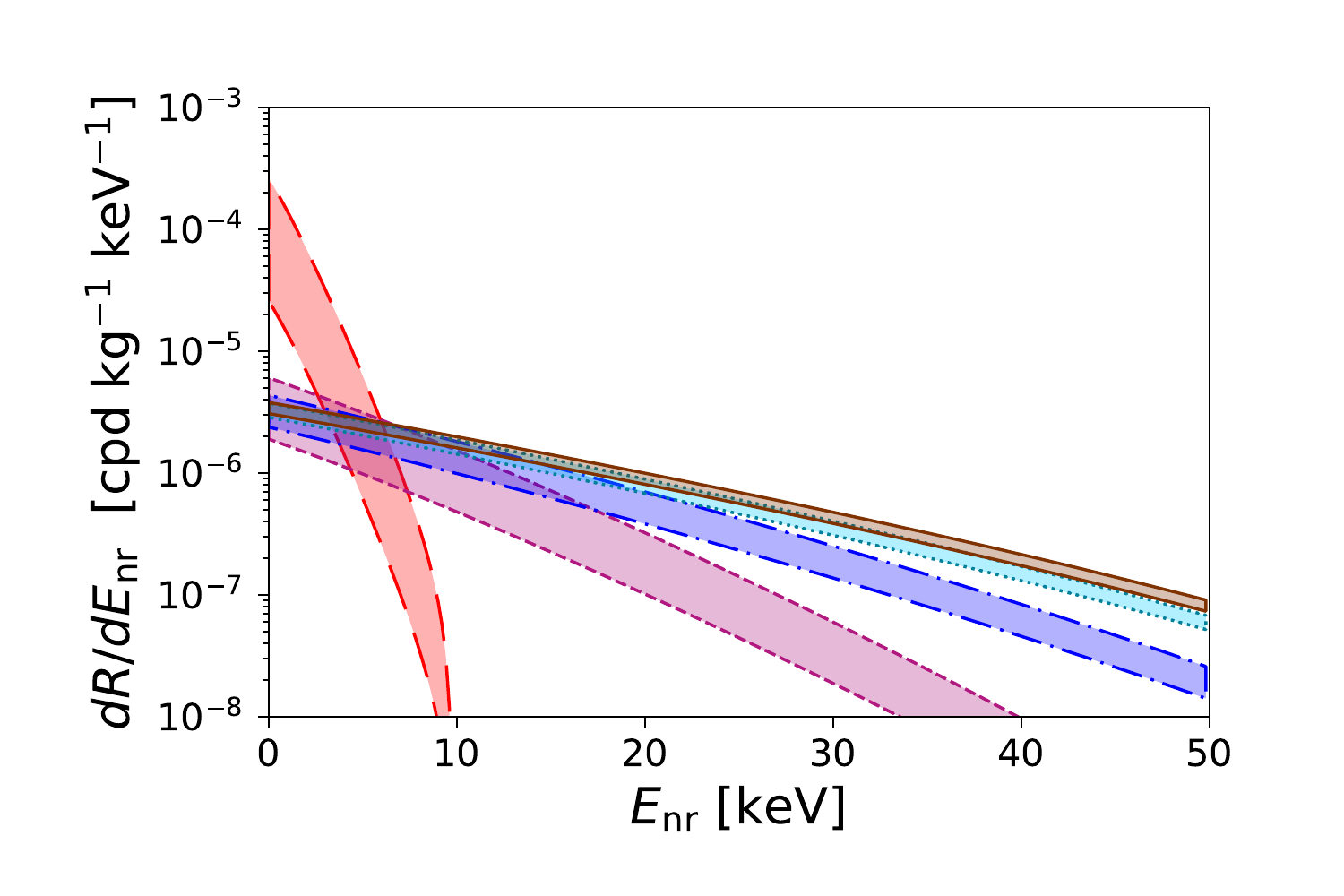}
\caption{Spin-independent differential recoil rate as a function of the nuclear recoil energy. The different edge line styles correspond to different neutralino masses, long-dashed: 10 GeV, short-dashed: 30 GeV, dotted-dashed: 50 GeV, dotted: 80 GeV and solid: 100 GeV. The area between lines with same neutralino mass indicates the allowed variation of the parameter $\mu$.}
\label{rate_vs_enr}
\end{figure}

In Figure~\ref{rate_vs_enr} we show the spin-independent differential direct detection rate (see Eq. \eqref{rate2}) as a function of the nuclear recoil energy for different WIMP's masses ($m_{\chi}=$10, 30, 50, 80 and 100 GeV) for June 2, where the laboratory speed is maximum. The width of each curve is related to the different values that the $\mu$ parameter can take for each mass of the WIMP. The greatest interaction between the WIMPs and the nucleus occurs at low nuclear recoil energy.

\subsection{SN neutrino signal analysis}\label{SNanalysis}

We have calculated the neutrino fluxes that arrive at the detector from a SN located at 10 kpc of distance, following Eqs.~\eqref{eq:ini-fluxes} and \eqref{flujos-3+1} for the {\it standard} case, and the $3+1$-scheme respectively. The signal was integrated over the first 10 seconds after the core bounce. Regarding the calculation for the active-sterile case, we considered the values shown in Table~\ref{dataactivos} for the active sector, and for the active-sterile mixing parameters, we have adopted the mixing values established by recent works \cite{Boser:2020, Diaz:2019, Dentler:2018}, that is $U_{e4}\sim 0.1$ and $\Delta m^2_{41}=1.3 \,\rm{eV}^2$. As stated previously (see Eq.\eqref{eq:PL}) we have considered a hindrance factor $h_s$ for the case of sterile neutrinos, to describe its presence at $t_{pb}=0$ inside the SN. For $h_s=0$, there is no sterile neutrino present in the neutrinosphere ($F_{\nu_s}^0=0$). The presence of initial sterile neutrinos in the SN is then indicated by taken $h_s\ne 0$, thus modifying the fluxes \cite{Tang:2020,Esmaili:2014} (see Eqs. \eqref{flujos-3+1}). 
\begin{table}[h!]
\caption{Active-active neutrino mixing parameters \cite{Pdg:2019}.} 
\begin{center}

{\renewcommand{\arraystretch}{1.3}
\begin{tabular}{ccc}
\hline\hline
Parameter & Normal hierarchy & Inverse hierarchy \\ \hline
$\sin^2(\theta_{12})$ & $0.307$ & $0.307$\\ 
$\Delta m^2_{21}$ & $7.53 \times 10^{-5} \, {\rm eV}^2$ & $7.53 \times 10^{-5} \, {\rm eV}^2$\\ 
$\sin^2(\theta_{23})$ & $0.545$ & $0.547$\\ 
$\Delta m^2_{32}$ & $2.46 \times 10^{-3} \, {\rm eV}^2$ & $-2.53 \times 10^{-3} \, {\rm eV}^2$\\ 
$\sin^2(\theta_{13})$ & $0.0218$ & $0.0218$\\
\hline\hline
\end{tabular}}
\label{dataactivos}
\end{center}
\end{table}

\begin{figure}[ht!]
\begin{center}
\includegraphics[width=0.8\textwidth]{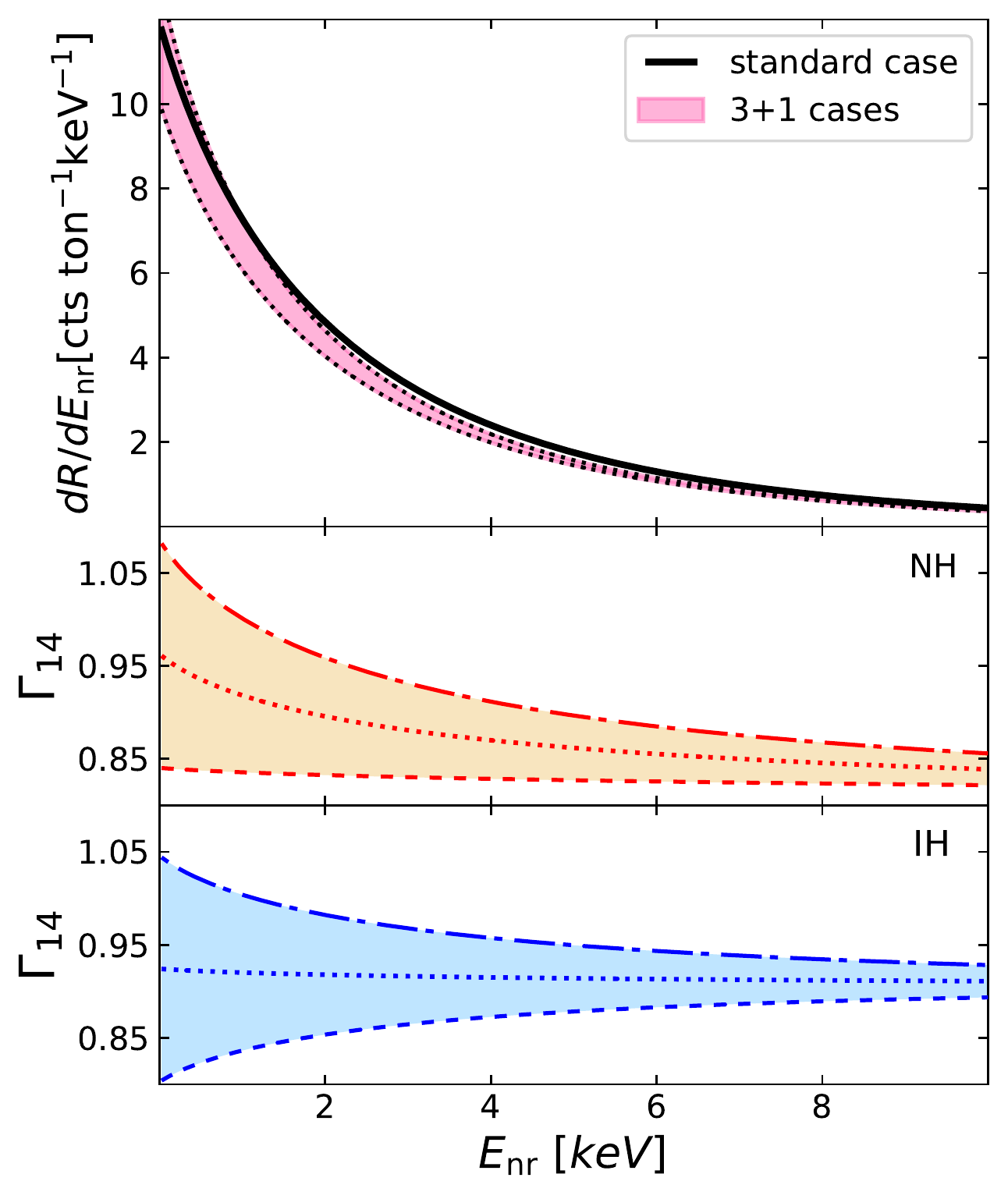}
\end{center}
\caption{Top panel: Neutrino differential scattering rates as a function of the nuclear recoil energy, with and without neutrino oscillations. Solid line: {\it standard} case (all flavors). The area between the dotted lines are the results obtained in the 3+1 scheme with $0\leq h_s \leq 1$. Middle and lower panels shows the results for the ratio $\Gamma_{14}=\frac{ dR/dE_{{\rm nr}_{3+1}}}{dR/dE_{{\rm nr}_{{\it standard}}}}$ for the normal (middle) and inverted (bottom) hierarchies. In both panels the upper dashed-dotted lines and the lower dashed-lines indicates the limiting values $h_s=1$ and $h_s=0$, respectively. The dotted lines are the results for $h_s=0.5$.}
\label{fig:dRdEr-3+1-alpha-cocientes-bestfit}
\end{figure}

We have calculated the differential recoil rates as a function of the recoil energy, for $N_{Xe}=4.6\times 10^{17} \, \rm{ton^{-1}}$. The top panel of Figure~\ref{fig:dRdEr-3+1-alpha-cocientes-bestfit} shows the calculated event rates with active-sterile neutrino oscillations, for the normal hierarchy (NH) and the inverse hierarchy (IH), along with the {\it standard} case. The shaded region of the upper panel stands for the rates obtained in the 3+1 scheme with $0\leq h_s\leq 1$.

To compare the effects due to the a sterile neutrino, we show the ratio $\Gamma_{14}=\frac{dR/dE_{{\rm nr}_{3+1}}}{dR/dE_{{\rm nr}_{{\it standard}}}}$ (bottom panel of Figure~\ref{fig:dRdEr-3+1-alpha-cocientes-bestfit}) as a function of the nuclear recoil energy. As before, shaded regions correspond to different values of $h_s$. The presence of a sterile neutrino generates a decrease in the rates with respect to the {\it standard} case, and it amounts $\approx15\%$ variation for $h_s=0$. For $h_s=1$, the rates with oscillation are higher than those corresponding to the {\it standard} case if $E_{\rm nr} \leq 2 \, \rm{keV}$. 

\begin{figure}[ht!]
\begin{center}
\includegraphics[width=0.8\textwidth]{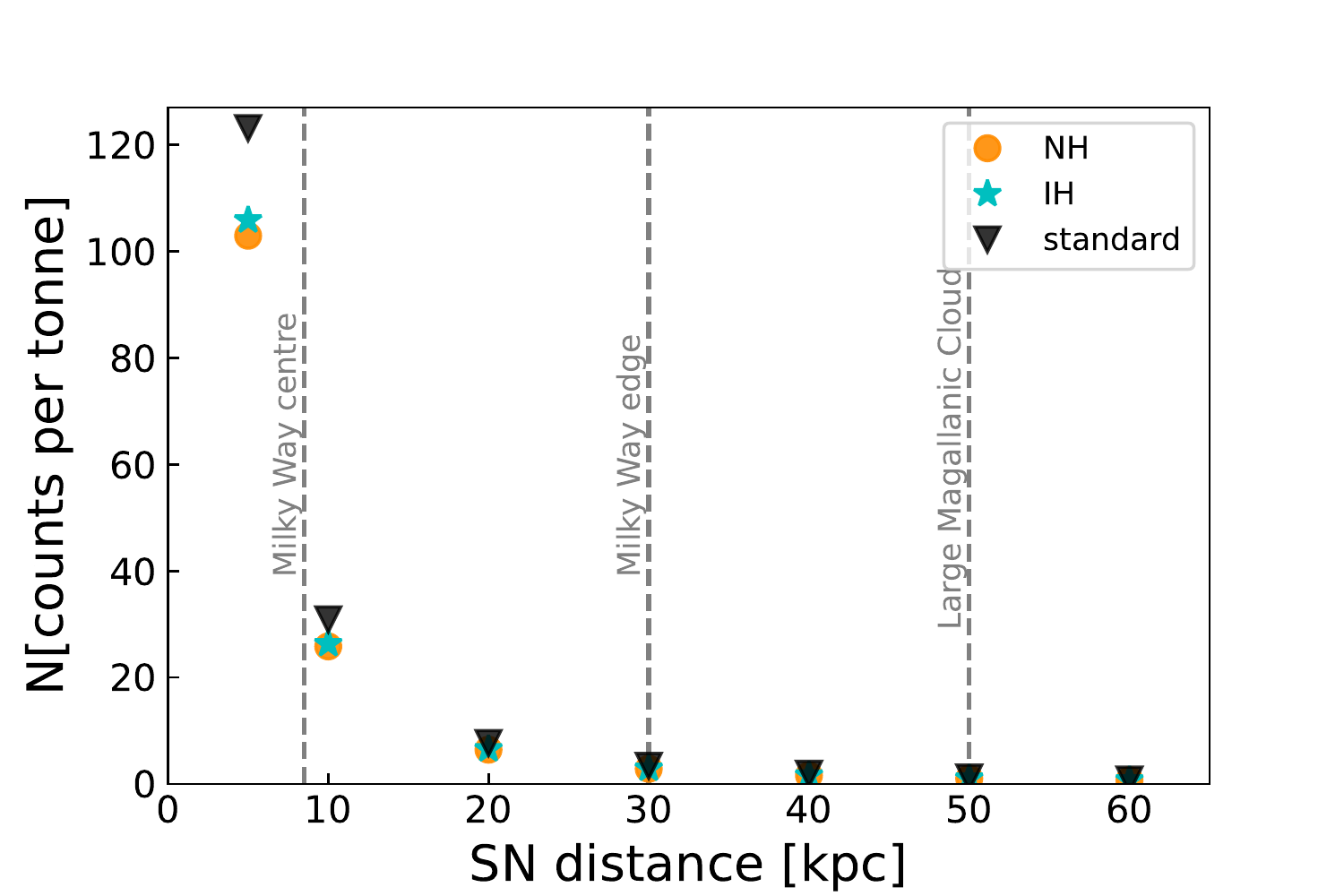}
\end{center}
\caption{Total counts per tonne of liquid xenon integrated on the recoil energy, 10 seconds before the bounce, as a function of the SN distance. } \label{fig:counts-vs-distance}
\end{figure}

Figure~\ref{fig:counts-vs-distance} shows the total number of counts as a function of the SN distance considering a perfectly efficient detector. For the active-sterile calculations, we show the results for $h_s=0$. As expected from the behavior of the rates, the case in the $3+1$ scheme produces lower counts than the {\it standard} case, being the case with NH the one for which the counts are the lowest, about $15\%$ with respect to the {\it standard} case. It is seen that for SN at distances greater than 20 kpc, it would be no longer possible to distinguish the counts related to the different cases.

\begin{figure}[ht!]
\begin{center}
\includegraphics[width=0.8\textwidth]{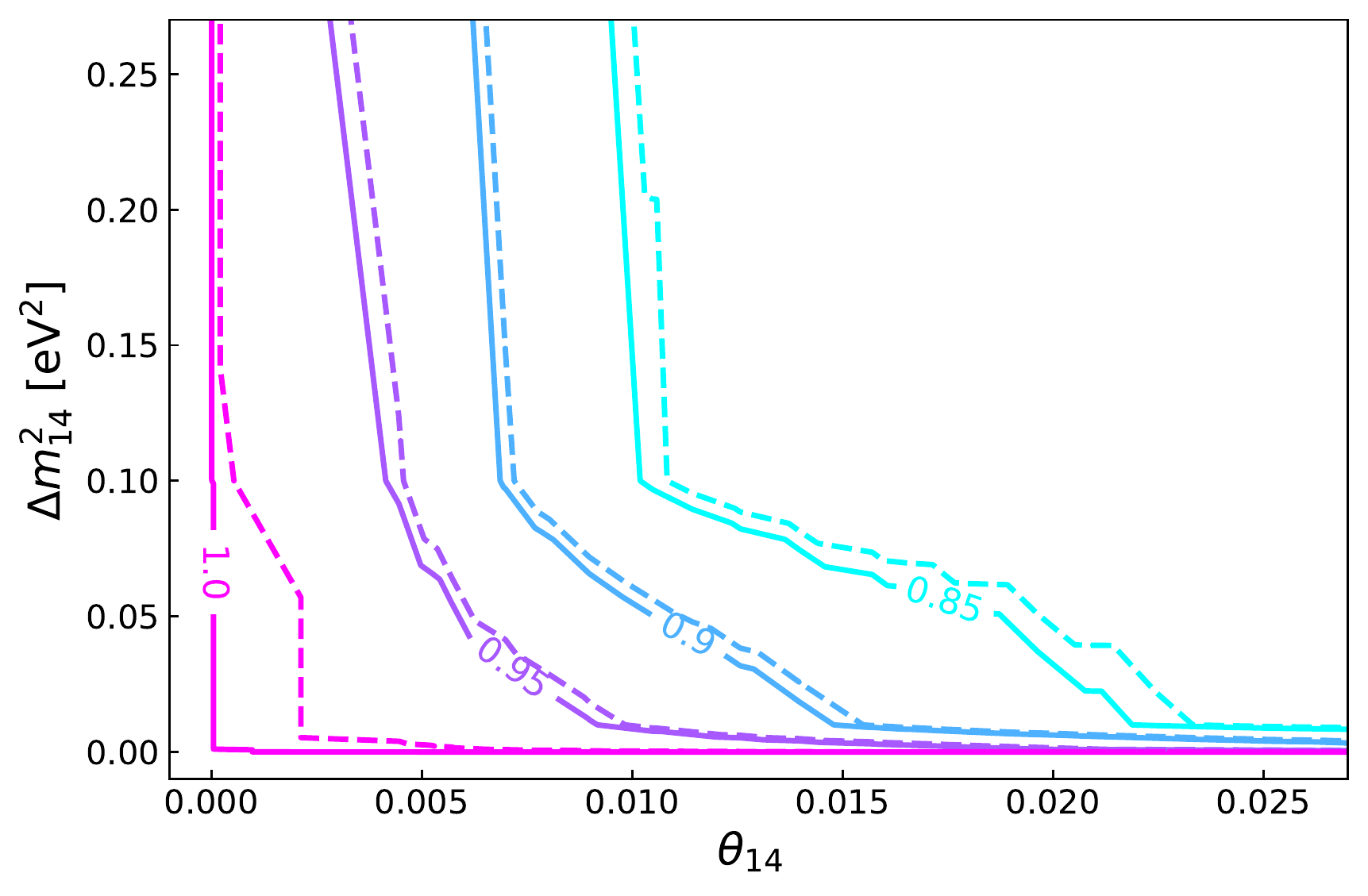}
\end{center}
\caption{Contour plots for the ratio $\frac{N_{3+1}}{N_{{\it standard}}}$ as a function of the active-sterile mixing parameters, with $h_s=0$. Solid lines: NH; dashed lines: IH. From left to right, the contours correspond to ratios of 1.0, 0.95, 0.9 and 0.85 respectively. } \label{fig:contornos_ef1}
\end{figure}

Then, we have studied the effects upon the signal produced by different values of the active-sterile mixing parameters. In particular we focus our attention to the ranges  $\Delta m^2_{41}\leq5 \, {\rm eV}^2$, $0\leq\theta_{14}\le \pi/4$ and $h_s=0$. In Figure~\ref{fig:contornos_ef1} we show the contour plots for the ratio between the number of events for the 3+1 scheme and the {\it standard} case ($\frac{N_{3+1}}{N_{{\it standard}}}$) as a function of the active-sterile mixing parameters for a perfectly efficient detector. The solid lines stands for the NH case and the dashed for the IH. 
For small values of active-sterile neutrino mixing parameters, one could expect similar counts than the {\it standard} case since the $P_S$ resonance is non-adiabatic (1.0 contours). For large values of the mixing parameters the counts are reduced up to a factor $0.85$. For real detectors, that is with a non-null threshold and a certain efficiency, the counts are slightly modified.

\subsection{Neutrino background for ANDES lab}
In order to set some predictions for ANDES laboratory, we have analyzed the dependence of the signal on the characteristics and location of the experiment (see Table~\ref{ANDES-parameters}). Different neutrino backgrounds, such as geoneutrinos, reactor neutrinos, solar and atmospheric neutrinos, could be measured by a dark matter detector. In particular, two location-dependent contributions are the backgrounds due to geoneutrinos and the reactor neutrinos. Geoneutrinos are electron antineutrinos originated from radioactive decays of $^{238}\rm{U}$, $^{232} \rm{Th}$, and $^{40} \rm{K}$ within the Earth’s interior whose fluxes are sensitive to the width of the Earth's crust under the lab. Its study can provide important information about the chemical composition of the Earth's interior and heat production mechanisms. KamLand and Borexino experiments have already reported geoneutrino observations through inverse beta decay reactions \cite{Araki:2005,Bellini:2010}. Due to its low intensity, the geoneutrino signal has not been intensively studied in direct searches, with the exception of some works \cite{Monroe:2007,Gelmini:2019}. On the other hand, the electron antineutrinos are also produced in nuclear reactors by fission beta decays of $^{238}\rm{U}$, $^{235}\rm{U}$, $^{239}\rm{Pu}$, and $^{241}\rm{Pu}$. The reactor neutrinos are the main source of background-noise, whose intensity depends on the location of the reactors. 

To calculate the geoneutrino signal, presented in Figure~\ref{nufloor}, we followed the References \cite{Machado:2012,Wan:2017,Huang:2013} and assumed a fully radiogenic Earth. 
The expected geoneutrino fluxes are shown in Table~\ref{geo_fluxes}. The reactor signal for ANDES laboratory was calculated taking into account the Argentinian and Brazilian reactors listed in Table~\ref{reactores}, and following the References \cite{Mueller:2011,Gelmini:2019}. The fission fraction and average released energy for each nuclear reactor isotope were taken from Reference \cite{Gelmini:2019} and for the neutrino spectra, we used the one modelled in Reference \cite{Mueller:2011}, based on a phenomenological fit to data.
\begin{table}[ht!]
\caption{Geoneutrino fluxes at ANDES lab.}
\begin{center}
{\renewcommand{\arraystretch}{1.2}
\begin{tabular}{cc}
\hline\hline
Component & Flux [$10^6$ cm$^{-2}$s${^-1}$] \\ 
\hline 
U&5.40 \\
Th & 5.05  \\
K& 24.04 \\
 \hline\hline
\end{tabular}}
\end{center}

\label{geo_fluxes}
\end{table}

\begin{table}[ht!]
\caption{Argentina and Brazil nuclear reactors near to the Andes laboratory.}
\begin{center}
{\renewcommand{\arraystretch}{1.2}
\begin{tabular}{cccc}
\hline\hline
Reactors & Power [MWt] & Location & Distance [km]\\
\hline
Atucha I & 1179 & $33^\circ 58'S$ $59^\circ 12'W$ &1084 \\
Atucha II & 2160 &  $33^\circ 58'S$ $59^\circ 12'W$&1084  \\
Embalse & 2064 & $32^\circ 13'S$ $64^\circ 26'W$&553\\
Angra I & 1882 & $23^\circ 0'S$ $44^\circ 27'W$ & 2640\\
Angra II & 3764 & $23^\circ 0'S$ $44^\circ 27'W$ & 2640\\
 \hline\hline
\end{tabular}}
\end{center}

\label{reactores}
\end{table}
The neutrino floor expected in the ANDES laboratory (shown in Figure~\ref{nufloor}) was constructed taking into account solar, atmospheric, reactor, and geoneutrinos interactions.  

\subsection{Dark matter and SN signal in the ANDES reference detector}

\begin{figure}[h!]
\centering
\includegraphics[width=0.8\textwidth]{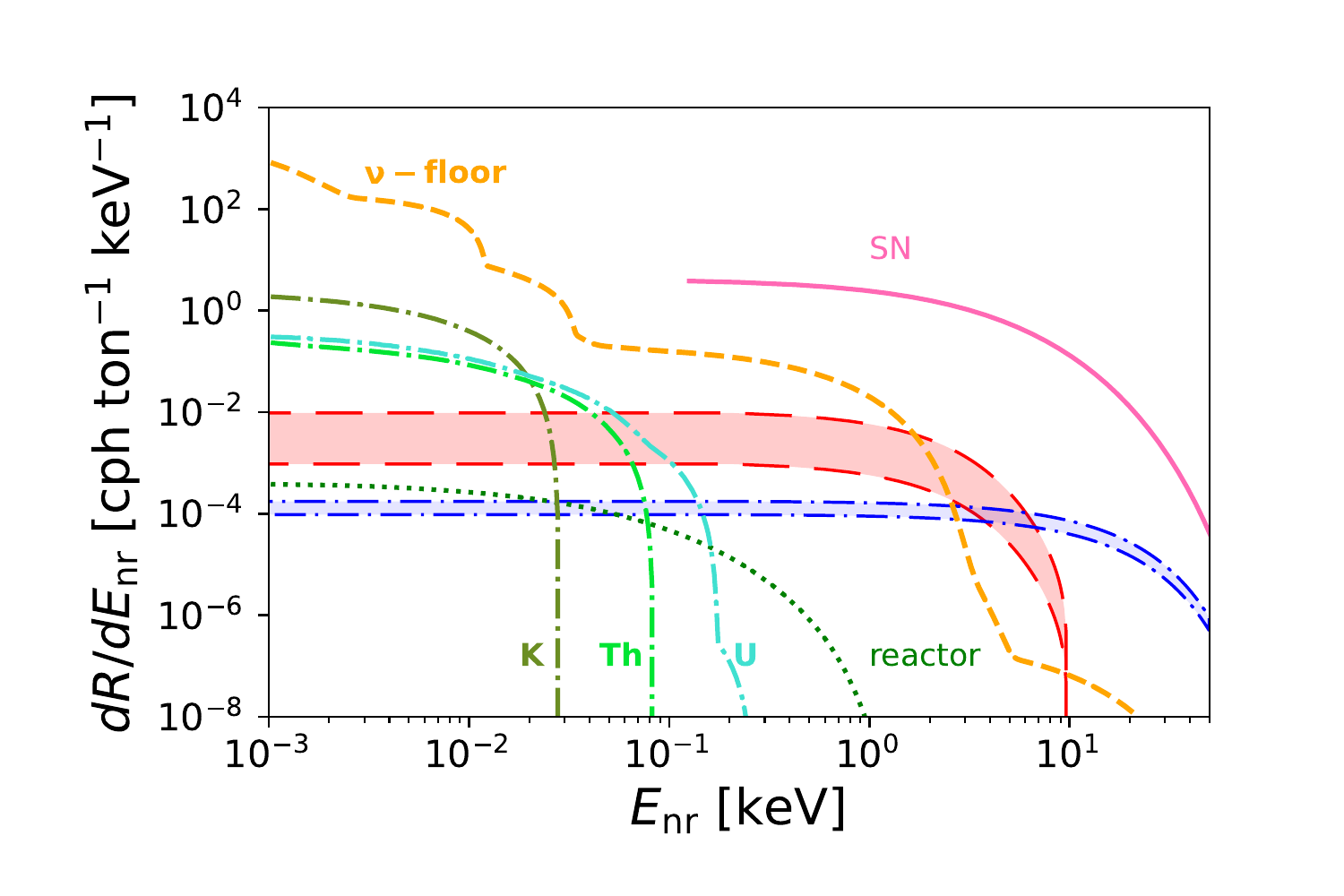}
\caption{Differential recoil rates for WIMPs (long-dash line: $m_{\chi}= 10$ GeV, dotted-dashed line: $m_{\chi}$=50 GeV) and SN neutrinos for the {\it standard} case (solid line) for the test $^{131}\rm{Xe}$ detector at the ANDES lab. The area between lines, for the WIMP's cases, denotes the allowed $\mu$ values. Neutrino background contributions for this site are also shown. Short dashed-line: total neutrino floor; dotted-line: reactor contributions; dotted-dashed-lines: geoneutrino contributions.}
\label{nufloor}
\end{figure}

In Figure~\ref{nufloor} we show the differential recoil rates expected for the ANDES laboratory as a function of the recoil energy for WIMPs and SN neutrinos. 
In this Figure, we show the results for two WIMP masses, $m_{\chi}=$10 GeV and $m_{\chi}=$50 GeV, the line width stands for different $\mu$ values. For the dark matter parameter space used in this work, we found that the WIMP signal starts to exceed the ANDES laboratory's neutrino floor for energies beyond 2.6 keV for $m_{\chi}>10$ GeV  (this energy could be around 0.6 keV if $m_{\chi}=7$ GeV). Thus, pursuing a detector with a lower energy threshold is not recommended unless we have a mechanism to disentangle the dark matter signal from the neutrino background.  At these scales, the cases with and without sterile neutrino in the neutrinosphere are practically indistinguishable.

\begin{figure}[ht!]
\centering
\includegraphics[width=0.8\textwidth]{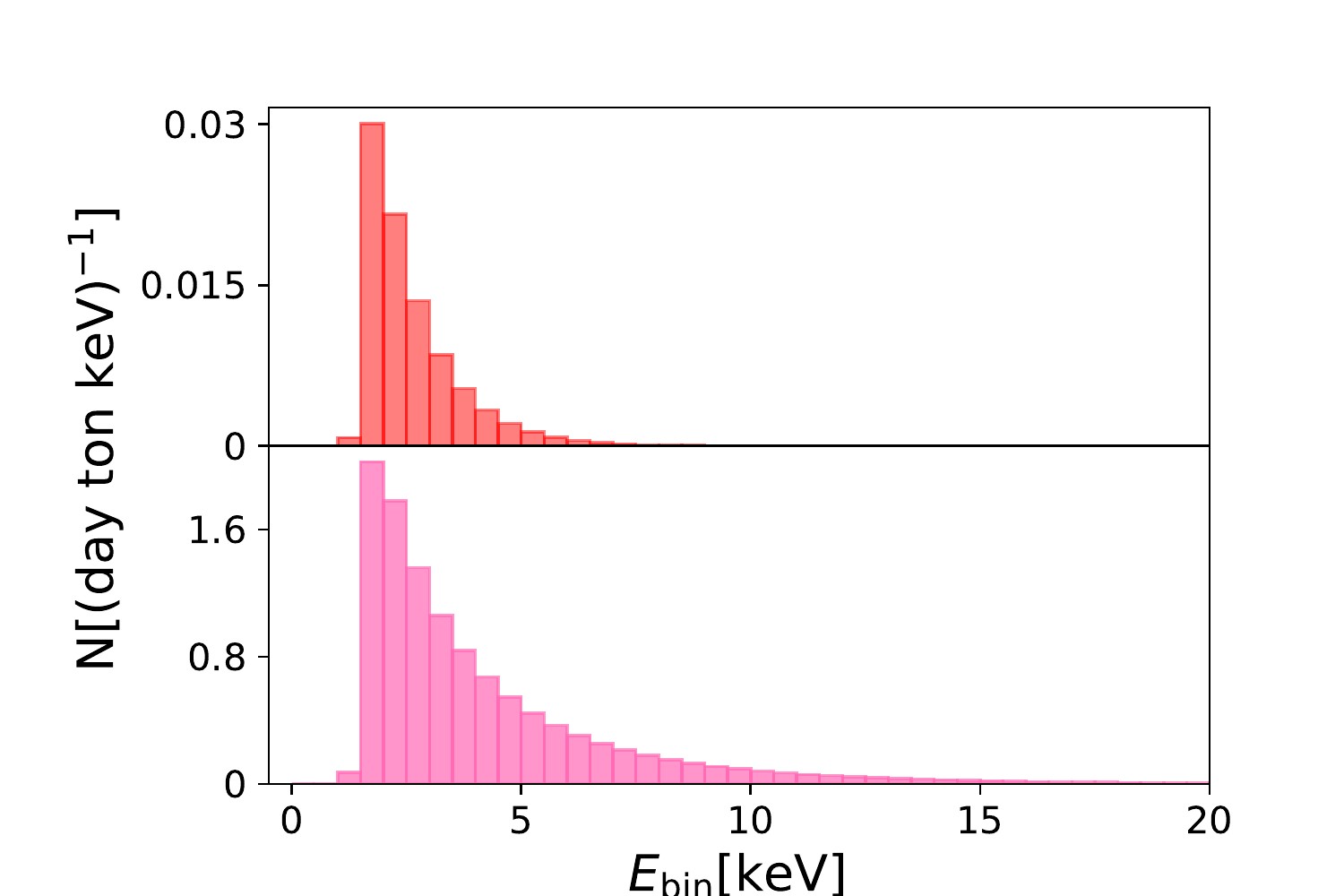}
\caption{Upper panel: WIMPs total counts in each energy bin, with mass of 10 GeV at the minimum value of the parameter $\mu$. \\ Lower panel: SN neutrino total counts for the {\it standard} case, in each energy bin for a SN located a 10 kpc of distance. }
\label{conteos_dm}
\end{figure}

To compute the total counts, we use Eq.~\eqref{counts_Xe1T} and the detector properties listed in Table~\ref{ANDES-parameters}.
The upper panel of Figure~\ref{conteos_dm} shows the counts distribution by bin for a WIMP of 10 GeV at the minimum value of the $\mu$ parameter, and the lower panel shows the SN counts distribution by bin for the {\it standard} case. Due to the efficiency and response function adopted, the first energy bins do not report counts in either case.

\begin{figure}[ht!]
\centering
\includegraphics[width=0.8\textwidth]{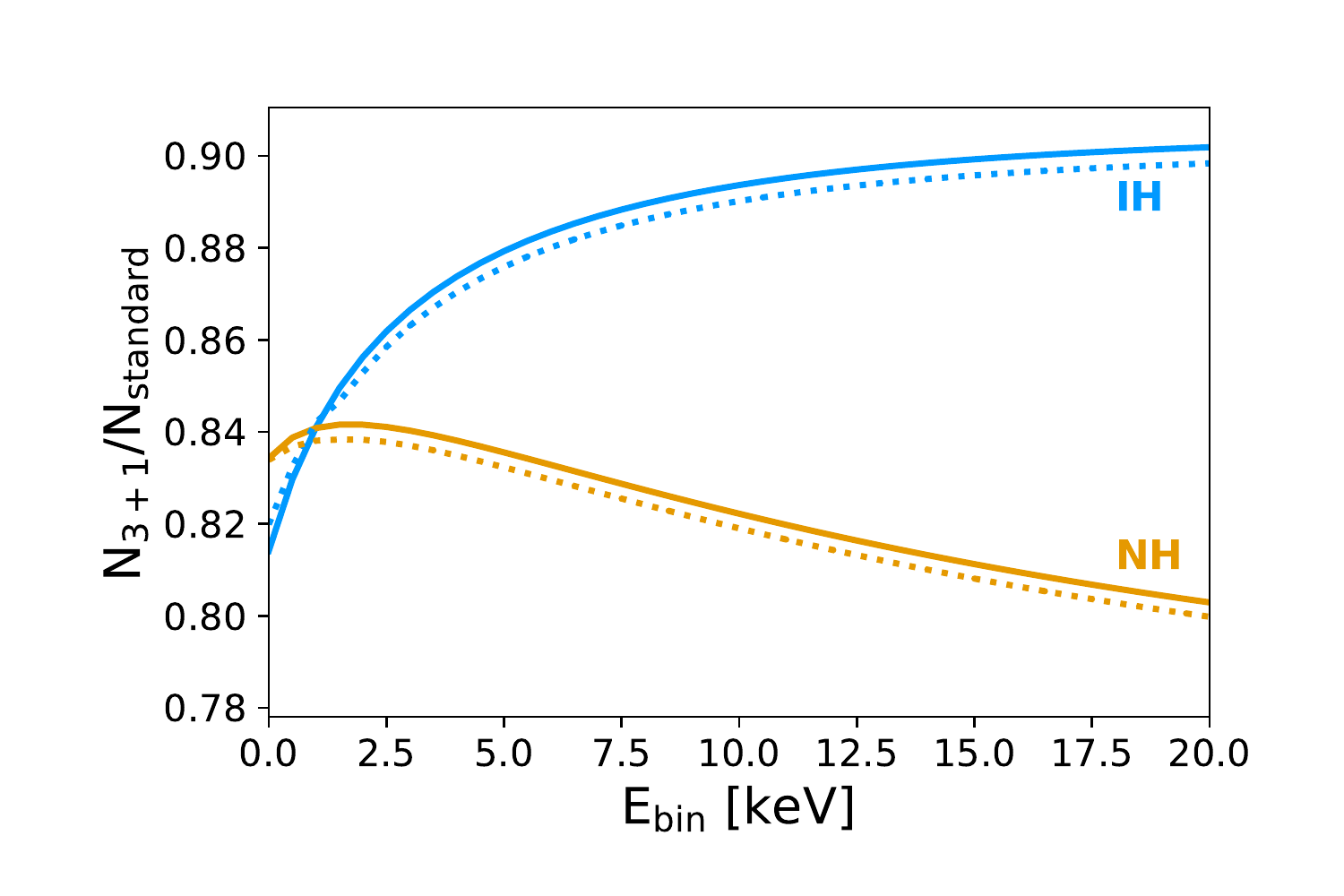}
\caption{Ratio between the counts for the active-sterile neutrino oscillation (3+1) and the {\it standard} cases. Solid lines show the results assuming perfect efficiency, and dotted lines assuming the actual Xenon1T efficiency. Both the normal (NH) and inverse (IH) hierarchies were considered.}
\label{cociente_hist_SN}
\end{figure}

We show in Figure~\ref{cociente_hist_SN} the ratio between the $3 + 1$ scheme (with $h_s=0$) and the {\it standard} case for each bin. Not remarkably differences between an ideal detector and our reference detector are observed for both the NH and IH schemes.

\begin{figure}[ht!]
\centering
\includegraphics[width=0.8\textwidth]{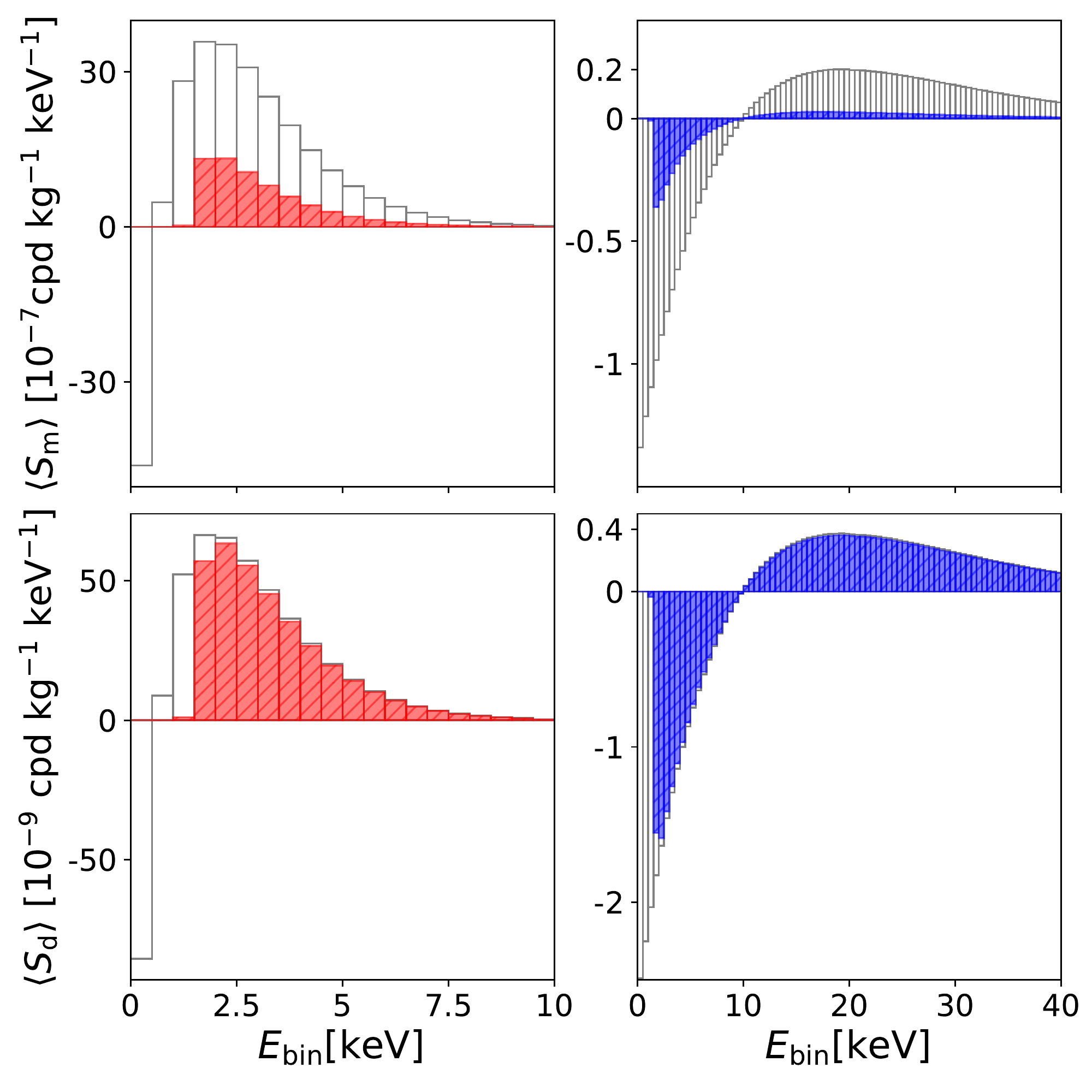}
\caption{The average annual (upper panel) and diurnal (lower panel) modulation amplitude as a function of the energy bin, at the minimum value of the parameter $\mu$. First column $m_{\chi}=10$ GeV, second column $m_{\chi}=50$ GeV. In all panels, the histograms show the results for an ideal detector and the hatched areas are those for the Xenon1T detector with the parameters given in Table~\ref{ANDES-parameters}.}
\label{anual_diurna_dm}
\end{figure}

As we mentioned before, annual modulation is a relevant feature for dark matter because most background signals are not expected to exhibit this time dependence. In Figure~\ref{anual_diurna_dm} we present the results for the average annual (upper panel) and diurnal (lower panel) modulation amplitudes by energy bin, for a WIMP of 10 GeV (left column) and 50 GeV (right column) at the minimum value of the parameter $\mu$. 
The shaded region, shows the average modulations considering an ideal detector, while the pattern region shows the results of considering the efficiency, dispersion and response function from Table~\ref{ANDES-parameters}. As we have shown for the total counts, the annual and diurnal modulation are suppressed in the first bins of energy when we considered a realistic detector. The diurnal modulation is two orders of magnitude lower than the annual modulation. The change of sign in the modulation for low energies occurs for an specific recoil energy that depends on the WIMP's mass, but it is independent of the value of the parameter $\mu$. The greater the mass of the WIMP the lower the modulation amplitude, and the sign change would occur at higher recoil energies. It can be seen that for lower WIMP masses (lower than 10 GeV) the change of sign would not occur for a detector with similar characteristics as Xenon1T, but this behaviour could be observed for a higher value of the mass of the WIMP, such as 50 GeV.

\subsection{Comparison of signals with Xenon1T}
Given that we consider a detector made of Xe with the efficiency and response function corresponding to the Xenon1T experiment, in this Section we analyze the expected variations in the signals due to different geographic locations (Andes and Gran Sasso Mountains). Among the signal studied in this work, those that depend on the Laboratory position on Earth are geoneutrinos, reactors, and diurnal modulation. 

\begin{figure}[ht!]
\centering
\includegraphics[width=0.8\columnwidth]{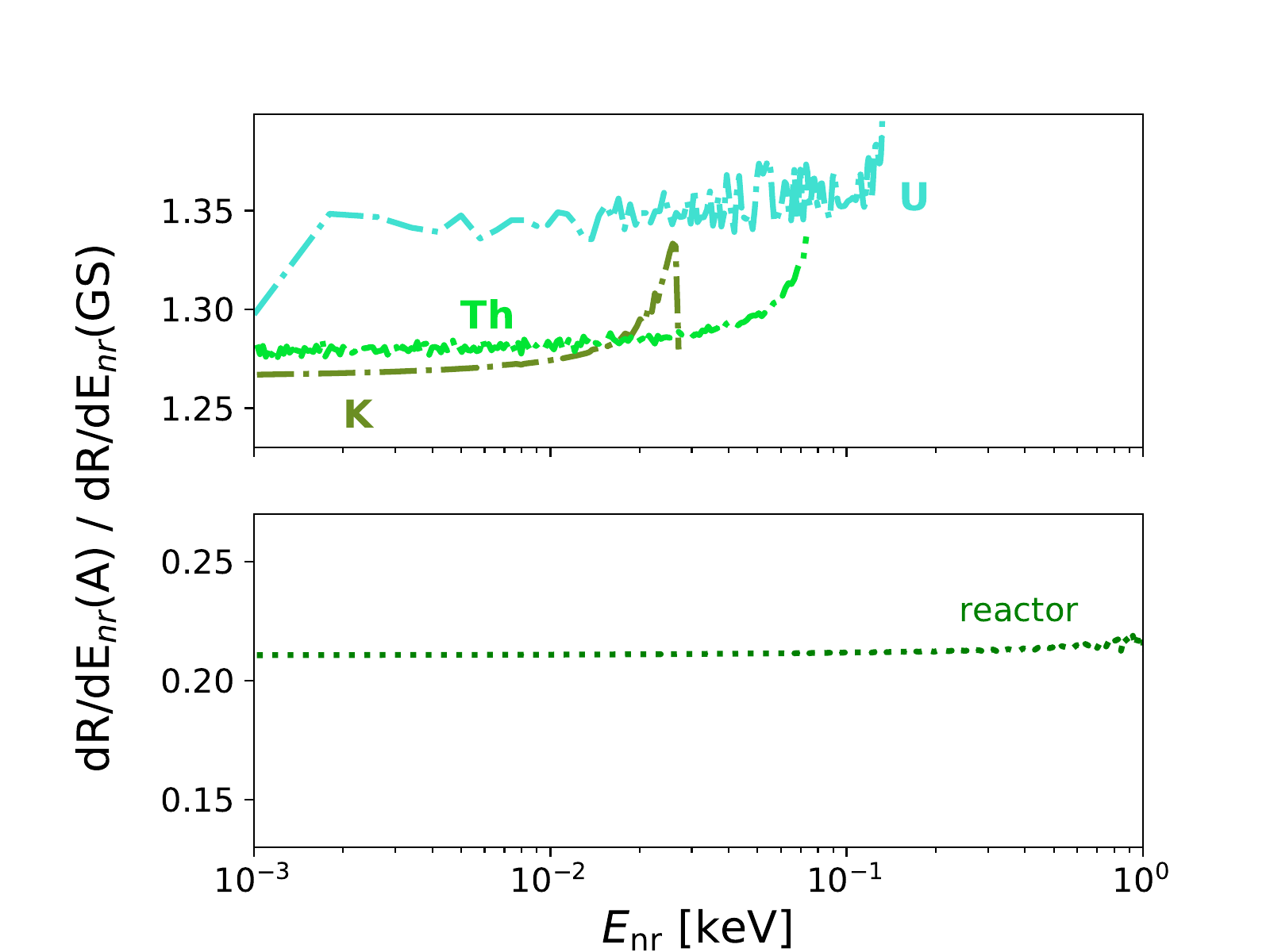}
\caption{Ratio of differential recoil rates (dR/dE$_{nr}$), for detectors located at the ANDES laboratory (A) and Gran Sasso (GS). Upper panel: geoneutrino signals. Bottom panel: reactor signals.}
\label{comparacion_ANDES_Xe1t}
\end{figure}

In Figure~\ref{comparacion_ANDES_Xe1t} we show the ratio between the expected rates in both ANDES and Gran Sasso locations for the geoneutrino (upper panel) and reactor (lower panel) components. 
 Given that the ANDES laboratory would be located in an area with one of the thicker Earth's crust and near the subduction of the Pacific and Continental tectonic plates, the geoneutrino signal might be relevant and significantly larger than in other laboratory sites \cite{Gelmini:2019, Huang:2013, Machado:2012}. In particular, we observe that the rate generated by geoneutrinos in ANDES is 20\% higher than expected in Gran Sasso. Instead, the reactor neutrino background in ANDES is expected to be 80\% smaller than the measured in Gran Sasso since the latter receives neutrinos from the Tricastin, Cruas, St. Alban, and Bugey reactors.
 
\begin{figure}[ht!]
\centering
\includegraphics[width=0.8\columnwidth]{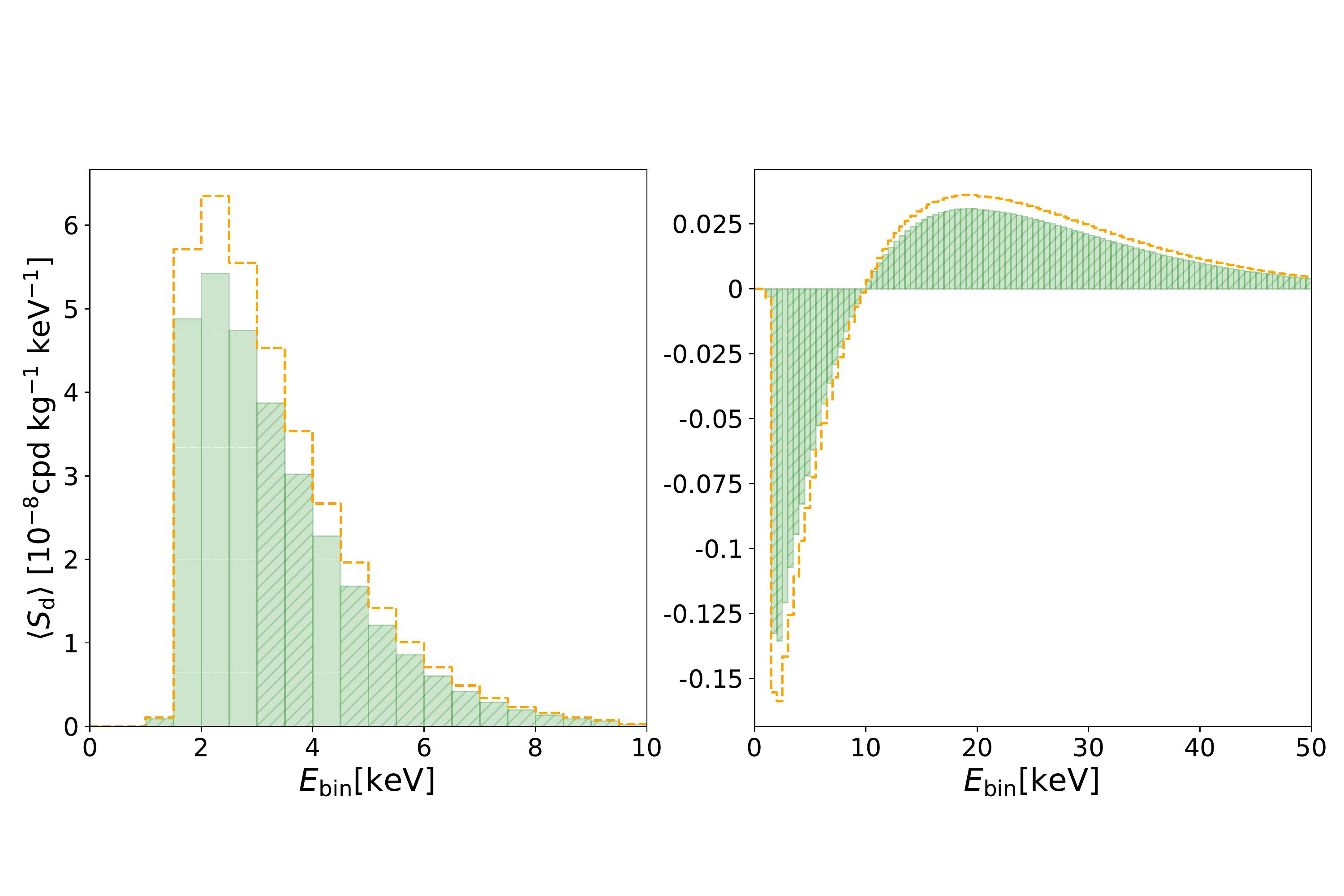}
\caption{The average diurnal modulation amplitude as a function of the energy bin, at the minimum value of the parameter $\mu$. Considering a real detector with the characteristics described in Table~\ref{ANDES-parameters}. First column $m_{\chi}=10$ GeV, second column $m_{\chi}=50$ GeV. The dashed line histograms show the results for ANDES, while the hatched areas stand for the Gran Sasso Xenon1T detector. }
\label{diurna_dm_andes_xenon}
\end{figure}

Regarding the diurnal modulation phenomenon, in Figure~\ref{diurna_dm_andes_xenon} we compare the amplitude that could be measured in ANDES and Gran Sasso experiments, for a detector similar to Xenon1T. It is observed a $17\%$ enhancement of the amplitude at ANDES respect to the one at Gran Sasso, which correlates with the latitude’s cosine of both sites. 

\section{Conclusions}\label{conclusion}

In this work, we have calculated the expected WIMP signal in direct detection of dark matter for the spin-independent case, considering the current limits established by Xenon1T and the neutrino floor. In addition, we have calculated the expected SN neutrino fluxes, including massive neutrinos, neutrino oscillations, and a light sterile neutrino. We also explored the possibility of the presence of sterile neutrinos inside the neutrinosphere. Furthermore, we studied the SN neutrino fluxes along the neutrino mixing parameter space and their respective differential scattering rates and counts.

We made specific predictions for the ANDES laboratory, considering a reference detector similar to Xenon1T (Gran Sasso). We evaluated the neutrino floor, including the neutrino fluxes from reactors and geoneutrinos specifics for the laboratory site. Finally, we have calculated the expected annual and diurnal modulation in our reference detector. 

We found that the expected recoil rate for different WIMP masses was most significant at low nuclear recoil energies.  For the SN signal, we found differences in rates and counts between the {\it standard} case and the 3+1 scheme of about 15\%, the case with normal hierarchy and $h_s=0$ was the one that reduces the counts the most. However, the 3+1 case in the non-adiabatic regime (small square-mass-differences and angles) could generate the same expected number of counts as the {\it standard} case. If a supernova explosion occurs, SN neutrinos would be responsible for 98\% of the counts reported within that day.

For the neutrino floor calculated for the ANDES laboratory, we found that the flux of geoneutrinos of K, Th, and U was relatively large compared to those expected in other locations. Also, we found that the dark matter recoil rates were above the neutrino floor for $E_{nr} \ge 3 \, \rm{keV}$. 

For the annual and diurnal modulation, we found that this effect would be more than five orders of magnitude lower than the total counts for our reference detector. But if the change of sign in the modulation amplitude is observed, it could bring valuable information about the mass of the dark matter particle. Finally, we compared the expected signal between ANDES and Xenon1T detectors to infer the features that should be observed due to the change of location.
We hope that these studies  might contribute to dark matter detection strategies that maximize the future ANDES laboratory detection capabilities.
\section*{Acknowledgements}
This work was supported by a grant (PIP-2081) of the National Research Council of Argentina (CONICET), and by a research-grant (PICT 140492) of the National Agency for the Promotion of Science and Technology (ANPCYT) of Argentina. O. C. and M. E. M. are members of the Scientific Research Career of the CONICET, M. M. S. is a Post Doctoral fellow of the CONICET and K. J. F. is a Doctoral fellow of the CONICET.

\bibliographystyle{ws-ijmpe}
\bibliography{bibliografia}
\end{document}